\documentclass{aa}  
\usepackage{graphicx}
 \usepackage[latin1]{inputenc}
 \usepackage{natbib}
\bibpunct{(}{)}{;}{a}{}{,} 
\def\kms{~km~s$^{-1}$}

\def\cmmt{~cm$^{-3}$}
\def\cmmd{~cm$^{-2}$}

\def\lct{$4_{04}-3_{03}$}
\def\lcc{$5_{05}-4_{04}$}
\def\lsc{$6_{06}-5_{05}$}

\begin{document}
   \title{HNCO enhancement by shocks in the L1157 molecular outflow}

   \subtitle{}

   \author{N. J.  Rodr\'{\i}guez-Fern\'andez  \inst{1}   
          \and M. Tafalla \inst{2} 
          \and F. Gueth \inst{1}
          \and R. Bachiller \inst{2}  
     }

   \offprints{N. J. Rodr\'{\i}guez-Fern\'andez}

   \institute{IRAM, 300 rue de la Piscine, 38406 St. Martin d'Heres, France \\
    \email{rodriguez@iram.fr}
    \and Observatorio Astronomico Nacional, IGN, 
     calle Alfonso XII 3, 28014 Madrid, Spain
      }

   \date{Received January 5, 2010; accepted March 22, 2010}

 

  \abstract
   {The isocyanic acid (HNCO) presents an extended
   distribution in the 
   centers of the Milky Way and the spiral galaxy    IC342. 
   Based on the morphology of the emission and the HNCO abundance with
   respect to H$_2$,  
   several authors made the hypothesis that HNCO could be a good tracer 
   of interstellar shocks.}
   {Here we test this hypothesis by observing a well-known Galactic
   source where the chemistry is dominated by shocks.}
   {We have observed several transitions of HNCO towards L1157-mm and  
   two positions (B1 and B2) in the blue lobe of the molecular outflow.}
   {The HNCO line profiles exhibit the same characteristics of other well-known
   shock tracers like CH$_3$OH, H$_2$CO, SO or SO$_2$. 
   HNCO, together with SO$_2$ and OCS, 
   are the only  three molecules detected so far
   whose emission is much more intense in B2 than in B1, 
   making these species valuable probes of chemical differences along
   the outflow.
   The HNCO abundance with respect to H$_2$
   is 0.4-1.8\,10$^{-8}$ in B1 and 0.3-1\,10$^{-7}$
   in B2. These abundances are the highest  ever measured, and imply an
   increment with respect to L1157-mm of a factor up to 83,
   demonstrating that this molecule is actually a good shock tracer.}
   {Our results probe that shocks can actually produce the HNCO abundance
   measured in galactic nuclei and even higher ones.  
   We propose that the gas phase abundance
   of HNCO is due both to grain mantles erosion by the shock waves 
   and by neutral-neutral reactions in gas phase involving CN and O$_2$. 
   The observed anticorrelation of CN and HNCO fluxes supports this scenario.
   The observed similarities of the  HNCO emission  and the sulfured molecules
   may arise due to formation pathways involving also O$_2$.   }


\maketitle

%

\section{Introduction}

Interstellar {\it isocyanic acid} (HNCO) was first detected towards Sgr B2 \citep{Snyder72, Churchwell86, Kuan96}.
Since the first detection in this source, the molecule has been observed in other
hot cores around massive \citep{Blake87, MacDonald96} and low mass protostars \citep{vanDishoeck95, Bisschop08}.
It has also been detected in translucent clouds \citep{Turner99} and
in the dense regions of Galactic molecular clouds \citep{Jackson84, Zinchenko00}, including those in the Galactic center
\citep{Huttemeister93, Lindqvist95,  Dahmen97, Rizzo00, Minh05, Martin08}.
HNCO has also been detected in some extragalactic sources  
\citep{Nguyen91, Meier05, Martin09}. 
The isotopologue HCNO (fulminic acid) has recently been detected by 
\cite{Marcelino09}.

The emission in HNCO lines with $K_{-1} = 0$
\footnote{
HNCO is a planar, nearly linear, slightly asymmetric  prolate rotor.  
The notation for a given level is $J_{K_{-1}K_1}$.
The $K_{-1} = 0$ are usually excited thermally, while
the excited $K$ ladders ($K_{-1} > 0$) are probably excited by FIR
radiation \citep{Churchwell86}.
}
is clearly extended in the two Galactic nuclei where 
it has been mapped: IC342 \citep{Meier05} and the Milky Way \citep{Lindqvist95, Dahmen97}.
In the Galactic center the spatial distribution of the
HNCO emission is different from
 that of most other tracers.
In particular, there is a HNCO peak without an associated CO peak at Galactic longitude $l=1.65^\circ$ \citep{Dahmen97}.
This very special distribution suggests that HNCO can be an important tracer of some  physical  processes that are not well revealed by other molecules.
It is possible that the molecule is tracing shocks, which are thought to take 
place at this Galactic longitude due to the gas dynamics in the barred potential
of the Milky Way \citep{Rodriguez08}.
Indeed, in the Galactic center, the highest gas phase abundances  of 
SiO (a well-known shock tracer)  are measured in this region 
 \citep{Huttemeister98, Rodriguez06}.
Also on the basis of a special morphology of the HNCO emission in Sgr B2, 
\cite{Minh06} suggested that HNCO is enhanced by shocks.
Regarding IC342, the HNCO emission resembles that of CH$_3$OH. In particular,
HNCO  is detected not only in the nuclear ring but also in
the inner spiral arms or \emph{dustlanes}.
Since the dustlanes are thought  to be 
the locus of strong shocks in barred galaxies 
\citep[see for instance][and references therein]{Rodriguez08},
\cite{Meier05} suggested that HNCO, as CH$_3$OH, could trace large scale 
shocks.   

Nevertheless, the hypothesis that HNCO is a good shock tracer at the scale
of galaxies still needs to be probed 
since the HNCO emission has never been studied
in well-known Galactic templates of interstellar shocks. 
In order to better understand the excitation and the chemistry of this promising
molecule, we have observed the protostar L1157 and its 
associated molecular outflow.
This outflow presents the morphological signature of
shocks \citep[][]{Gueth98, Codella09} and
it is frequently used to benchmark numerical models of shocks 
\citep{Gusdorf08a, Gusdorf08b}. 
L1157 is the best example of
``chemically active outflow" \citep[][]{Bachiller01}
and a template of shock chemistry, 
since many species exhibit large abundance increments with respect to the
protostar \citep[][]{Bachiller97}.
Therefore,
it is a source of choice to characterize the emission of a given molecule
in a shocked environment 
\cite[see for instance][]{Bachiller97, Bachiller01, Benedettini07, 
Arce08, Codella09}.

\begin{figure}
\begin{center}
\includegraphics[angle=-90,width=8cm]{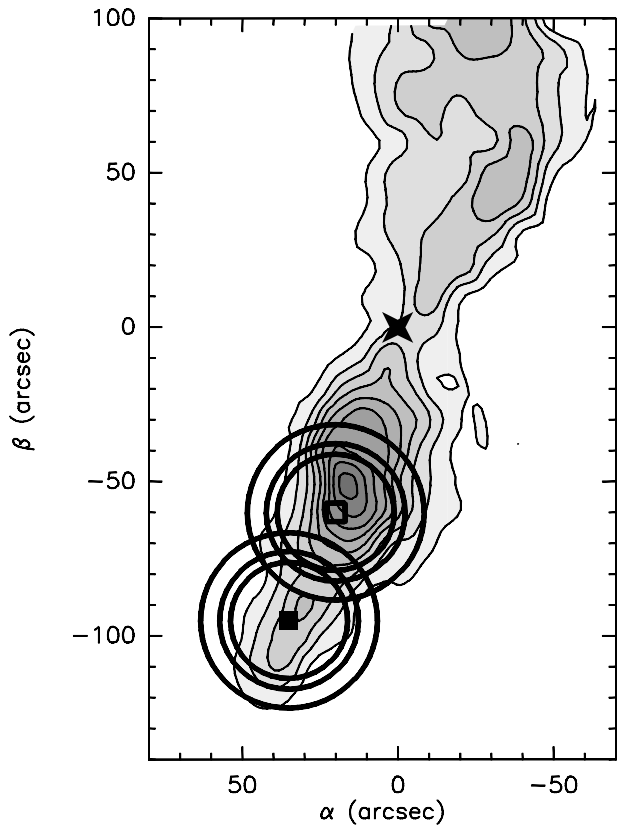}
\end{center}
\caption{ Integrated CO(2-1) emission from the L1157 outflow \citep{Bachiller01}.
We show the position of L1157-mm (star), B1 (empty square) and B2 (filled square).
B1 and B2 are the positions of the shocks as traced,  for instance, by SiO(3-2)
\cite[see Fig. 6 of][]{Bachiller01}.
The CO emission peaks behind the actual positions of the shocks.
The different circles represent the half maximum contours of the IRAM 30m
 beam at the frequencies of the lines discussed in this paper (see Table~\ref{tab:obs}). }
\label{fig:l1157}
\end{figure}

\section{Observations}
\label{sec:obsres}

We have observed three lines of the $K_{-1} = 0$ ladder of HNCO 
towards the two main shocks in the southern lobe of the L1157 molecular outflow:
the B1  and B2 positions of \cite{Bachiller97}.
In addition, we have  observed as reference the continuum source L1157-mm
(the protostar).
The equatorial coordinates of L1157-mm are RA=$20^h39^m06.19^s$ Dec=$68^\circ02'15.9''$, (J2000).
The offsets of B1 and B2 with respect to L1157-mm are
 ($20", -60"$) and ($35", -95"$), respectively.

The observations were done with the IRAM 30m telescope in Pico Veleta in July 2007.
The line quantum numbers, frequencies and the telescope parameters are listed in Table \ref{tab:obs}.
As backends we used the 100 KHz filterbank and
the VESPA autocorrelator with a channel resolution of 20-40 MHz,
which allows to study in detail the line wings.
The observations were carried out in position switching mode with the {\sc off}
position located 24' NE from  L1157-mm.
Typical system temperatures and rms noise of the spectra are 
are given in Table \ref{tab:obs}.

Table \ref{tab:lines} gives
the integrated flux of the different lines computed from spectra in 
 $T_{mb}$ units (forward and beam efficiencies are given in Table \ref{tab:obs}).
The uncertainties in the calibration of the IRAM 30m telescope are within 
10 $\%$  at the frequency of the lines discussed here \citep{Mauersberger89}.

\begin{table}
\caption{Observational parameters. }
\label{tab:obs}
\centering
\begin{tabular}{lllllll}
\hline
Line & $\nu$\tablefootmark{a} & B$_{eff}$\tablefootmark{b} & F$_{eff}$\tablefootmark{c} &  FWHM\tablefootmark{d}   & T$_{sys}$\tablefootmark{e} & rms \\
        & GHz   &                   &                  & arcsec& K  & mK \\
\hline
4-3\tablefootmark{f}    & 87.925    & 0.78  & 0.95 & 28.4 &100& 12-18\\
5-4\tablefootmark{g}     & 109.905   & 0.75  & 0.95 & 22.3 &180& 16-19 \\
6-5\tablefootmark{h}     & 131.885   & 0.69  & 0.93 & 18.8 &215& 28-34\\
\hline
\end{tabular}
\tablefoot{
\tablefoottext{a}{Frequency}
\tablefoottext{b}{Beam efficiency}
\tablefoottext{c}{Forward efficiency}
\tablefoottext{d}{Full width at half maximum of the telescope main beam}
\tablefoottext{e}{System temperature}
\tablefoottext{f}$4_{0,4}-3_{0,3}$
\tablefoottext{g}$5_{0,5}-4_{0,4}$
\tablefoottext{h}$6_{0,6}-5_{0,5}$ 
}
\end{table}

\section{Results}
\subsection{Line profiles: differences between the observed positions}
\label{sect:profiles}

\begin{figure*}
\begin{center}
\includegraphics[angle=-90,width=14cm]{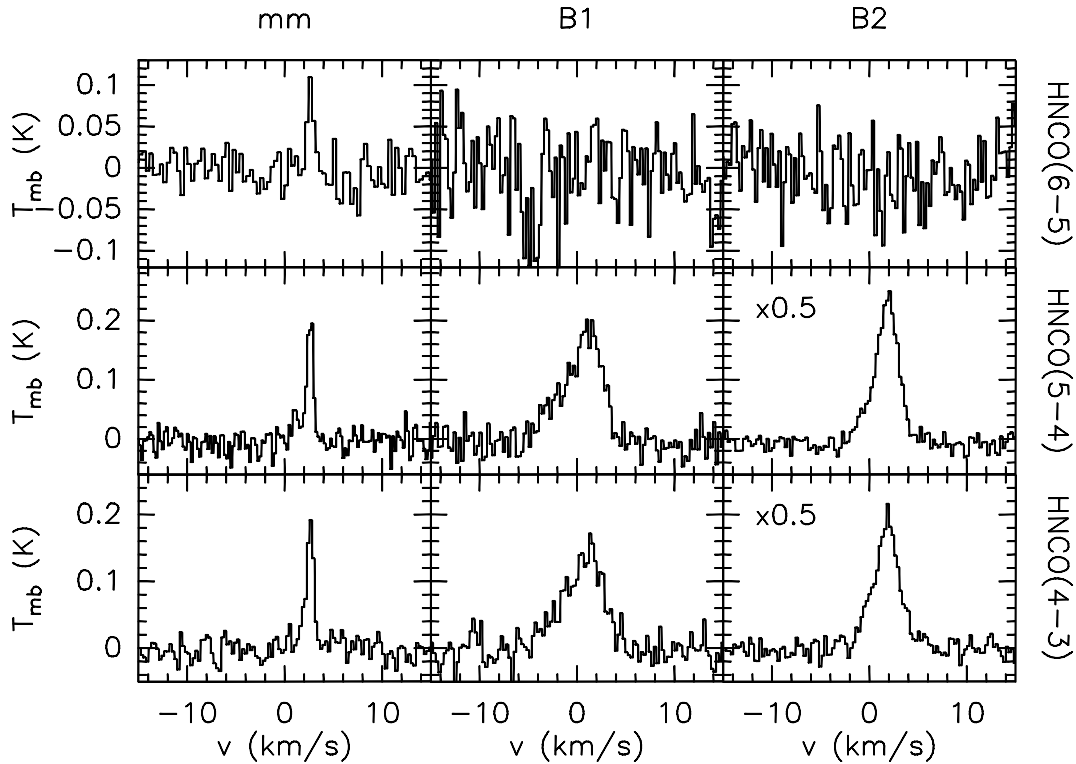}
\end{center}
\caption{Observed HNCO spectra. From bottom to top the different panels
show the HNCO \lct , \lcc \ and \lsc \  spectra
towards  L1157 mm, B1 and B2. }
\label{fig:hncolines}
\end{figure*}

Figure \ref{fig:hncolines} shows the spectra. 
The \lct\ and \lcc\ lines have been detected in all the positions.
In addition the \lsc \ line has also been detected toward L\,1157\,mm.
The lines are narrow  towards L1157-mm (linewidths of 0.7-0.9 \kms) 
but the linewidth increases
by more than a factor of 3 towards the B1 and B2 positions.
In particular, they show a prominent blue-wing towards B1.

\begin{figure}
\includegraphics[width=8cm]{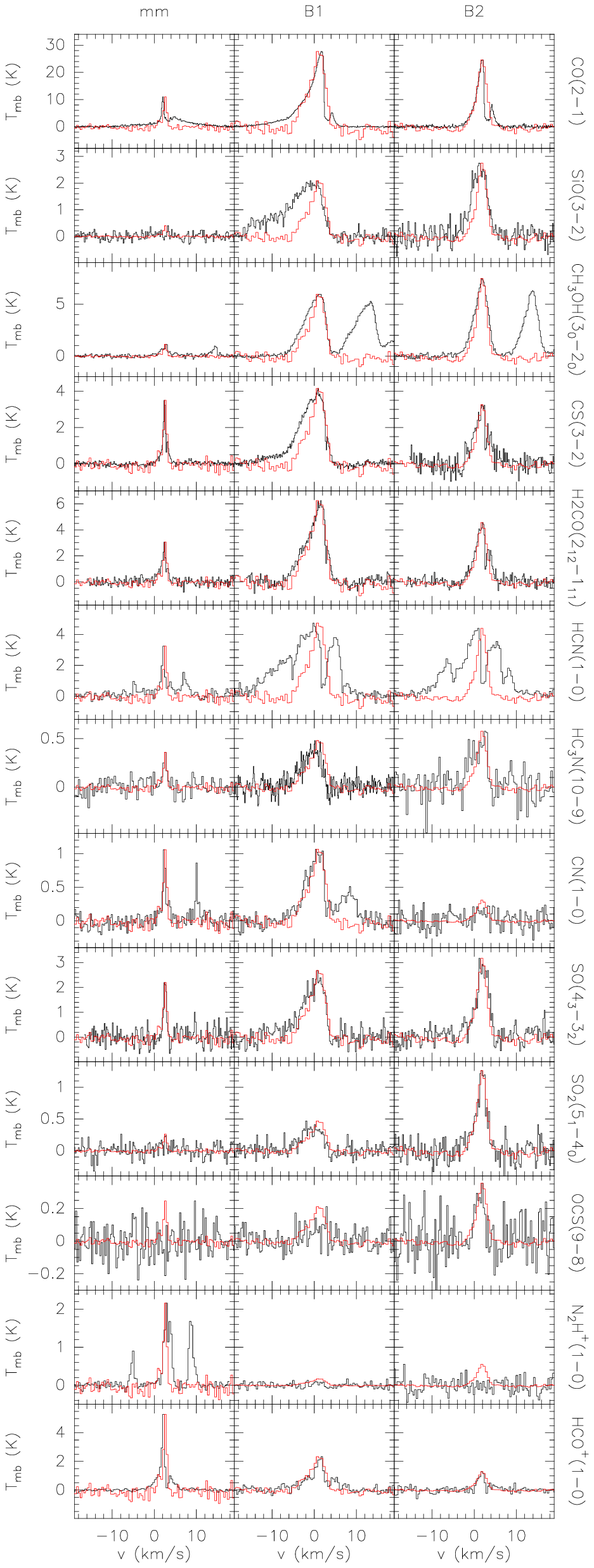}
\caption{Comparison of the HNCO(\lcc) line profiles (in red)  with the spectra of other species published by Bachiller \& Perez-Gutierrez (1997), in black lines.
The HNCO peak temperature has been scaled to the peak temperature of the other molecule in order to compare profiles.}
\label{fig:bpglines}
\end{figure}

Figure \ref{fig:bpglines} shows a comparison of the HNCO (\lcc) line profile
measured towards the three observed positions and the 
spectra of other molecules observed by \cite{Bachiller97}.
The line profiles are in overall good agreement with the other molecules.
In particular, the HNCO profiles are very similar 
to those of CH$_3$OH, H$_2$CO, SO and SO$_2$,
which are all 
 shock tracers whose abundance has increased by more than a factor of 10 in the shocked gas with respect to the quiescent gas in L1157-mm.
The HNCO line wings are narrower than SiO lines, in agreement with
 the findings  of \cite{Zinchenko00} in Galactic dense cores, but the SiO  line profiles are
specially broad in the L1157 molecular outflow, being broader than
those of all the other molecules as well.

As discussed by \cite{Bachiller97}, most of the molecules exhibit lines with
similar intensity in B2 and in B1. The exceptions are CN,
which is very intense towards B1 but it is not detected towards B2, and the 
sulfur-bearing  molecules SO$_2$ and OCS,
whose lines are much more intense in B2 than in B1.
It is interesting to point out that HNCO,
whose lines are a factor of 2 more intense in B2 than in B1, 
is indeed the only non sulfur-bearing 
molecule showing this behavior.

\subsection{Enhanced HNCO abundances in the shocked gas}
\label{sect:abundances}
The critical density of the HNCO transitions discussed here is high 
\citep[ 5\,10$^5$-10$^6$ \cmmt; derived using the collisional coefficients 
in the {\sc lamda} database for temperatures
from 20~K to 320~K,][]{Schoier05}.
Therefore, it is likely that most of the emission
comes from  dense and small clumps as those found
in B1 with the IRAM Plateau de Bure interferometer 
\citep{Benedettini07, Codella09}. 
For instance, the  angular size of the CH$_3$OH clumps is 10".
However, without high angular resolution observations it is
difficult to assign a very precise value to the size of the source emitting in HNCO.
Therefore, all the following results have been obtained
in two limiting cases:
first, assuming that the size of the emitting region is 10" and computing source
brightness temperatures taking into account the corresponding beam filling factor.
Second, 
assuming that the emitting region fills the primary beam  of the telescope,
and thus using $T_{mb}$ temperatures.

\subsubsection{LTE analysis}
We have computed HNCO column densities assuming optically thin emission and LTE excitation.
We have computed excitation temperatures from the $5_{05}$
and $4_{04}$ populations { (T$_{ex}^{54}$) and from the $6_{06}$
and $4_{04}$ populations (T$_{ex}^{64}$).
The temperatures T$_{ex}^{64}$ are very low (5-6 K for mm and $<6$ K for B1 and B2).
The excitation temperature  T$_{ex}^{54}$  varies from 
8 to 11 K for L1157-mm (depending on the assumptions on the source  size),
while it varies  from 9 to 19 K in B1 and B2 (see Table~\ref{tab:results}). 
In order to study with great precision the HNCO excitation more than three transitions would
be needed.
Here, in order to obtain a conservative estimation of the total column
density of HNCO we have
extrapolated the population in the  $5_{05}$ and $4_{04}$ levels
using the T$_{ex}^{54}$.
Of course, the total HNCO column density would be higher if the excitation
temperature is actually lower. }
Using T$_{ex}^{54}$  and assuming LTE excitation, 
the total HNCO column densities are
0.2-0.8, 1-1.9 and 1.5-3.8 (in units of $10^{13}$ \cmmd)
for mm, B1, and B2, respectively.
Since the excitation is similar in B1 and B2, the highest line
intensities measured 
towards B2 translate into larger column densities, in contrast to 
the CO column density
which is a factor of 2 lower in B2 \citep{Bachiller97}.

\subsubsection{{\sc radex} analysis}
In addition, we have  studied the HNCO excitation 
using  {\sc radex} \citep{vanderTak07},
which is a non-LTE excitation and radiative transfer code that
decouples the statistical equilibrium and the radiative transfer 
equations using the escape probability method.
The calculations assume lines of a given width and rectangular shape.
Therefore, the change of the optical depth over the profile is not taken
into account. This is not a problem if the opacity of the lines is
not very high.
We have used the HNCO-H$_2$ collisional coefficients listed in the
 {\sc lamda} database \citep{Schoier05}.
 The computations use a 
background temperature of 2.73 K. 
We have modeled the line fluxes (or integrated intensities) as a 
function of the hydrogen volume density ($n_{H_{2}}$), the 
HNCO column density ($N_{HNCO}$), the kinetic temperature ($T_K$) and
the velocity dispersion.
The range of hydrogen densities used is $10^3-10^6$ \cmmt \ and that of 
$N_{HNCO}$ is $10^{12}-10^{15}$ \cmmd.
Two kinetic temperatures have been considered:
 a low kinetic temperature (30 K), only slightly higher than
the excitation temperature derived with the HNCO lines and a
much higher temperature (250 K), since the actual gas temperature
in the shocked gas can be considerably higher than the HNCO excitation temperatures.
These temperatures probably cover the different physical conditions of L1157-mm
 and the shocked sources B1 and B2.
We have computed models 
with velocity dispersions (linewidths) from 1 to 6 \kms , which 
cover the measured range of linewidths in the three observed positions. 
In the range of physical conditions compatible with the observations, 
the opacity at the peak of the lines is low. 
Therefore, the predicted integrated intensities
 are independent of the actual
linewidth used to compute the models.
For instance, the figures discussed below have been computed 
with a linewidth of 3 \kms .

\begin{figure}
\begin{center}
\includegraphics[angle=-90,width=8.4cm]{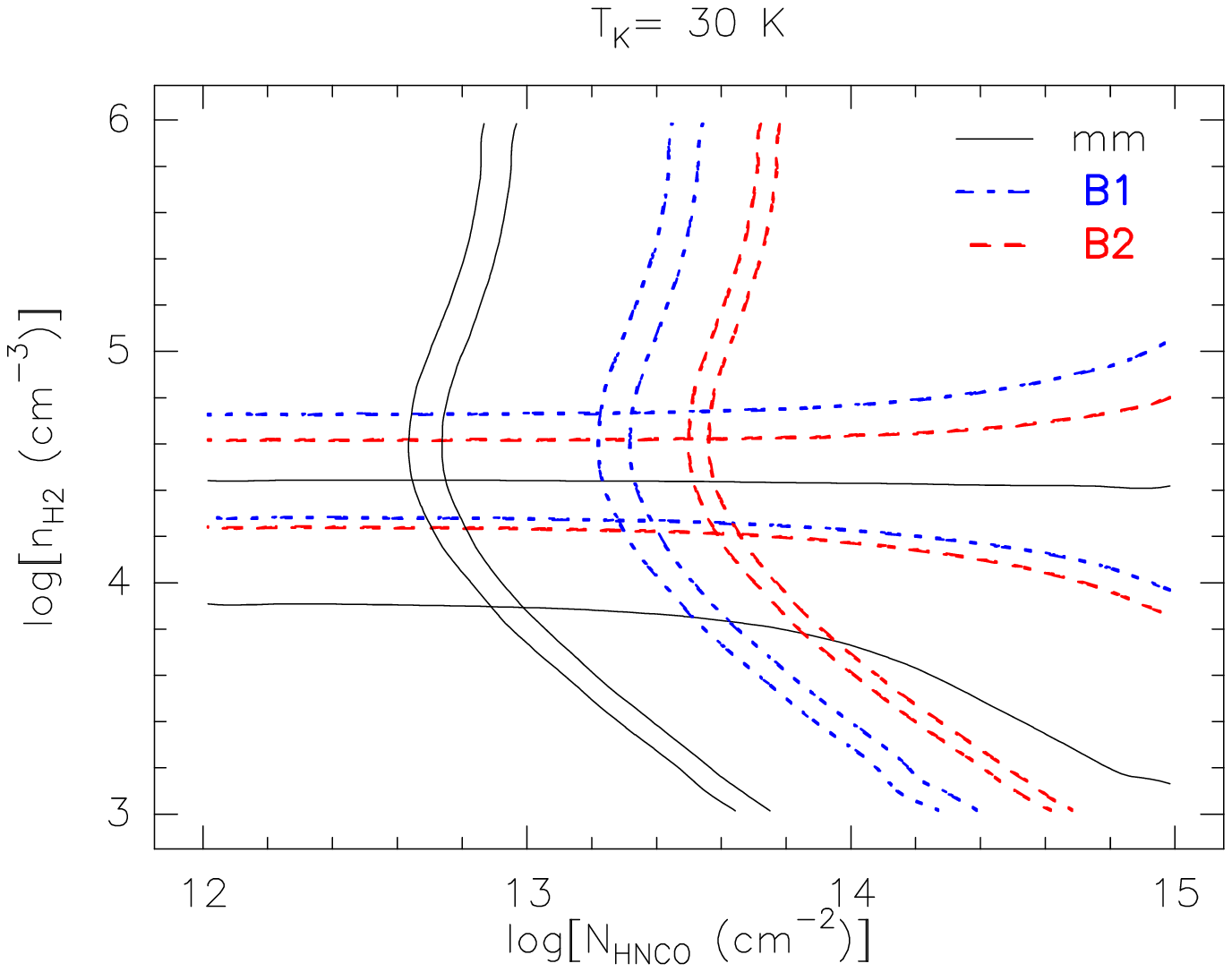}\\
\vspace{2mm}
\includegraphics[angle=-90,width=8.4cm]{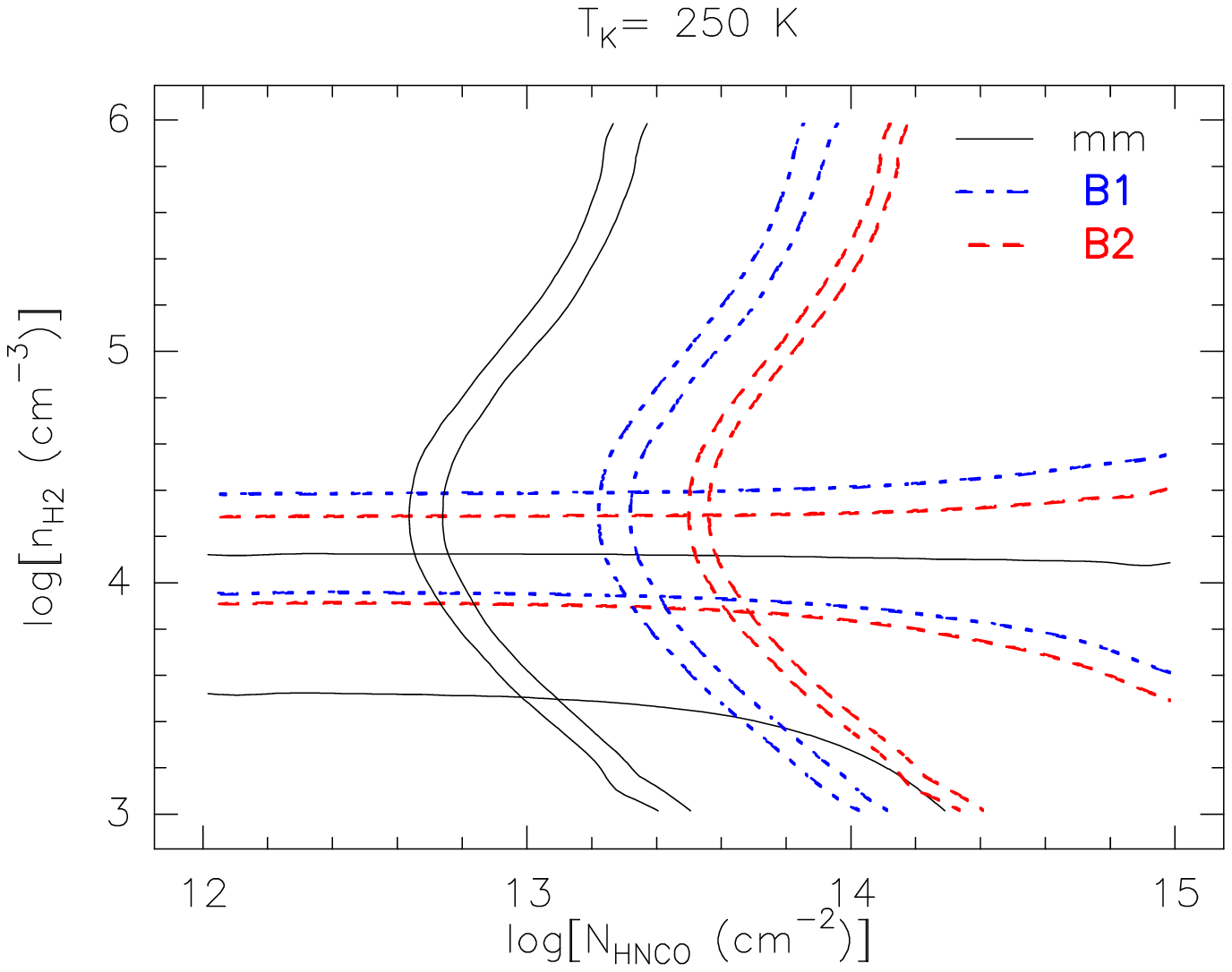}
\caption{{\sc radex} results for the $5_{05}-4_{04}$ to  $4_{04}-3_{03}$ 
flux ratio
(horizontal curves) and the $4_{04}-3_{03}$ line flux
(vertical curves) computed with a 
brigthness temperature for a source size of 10" 
as a function of the hydrogen density, the HNCO column density and the kinetic 
temperature (30 K in the upper panel and 250 K in the lower panel). A typical 
uncertainty of 10 $\%$ in the line flux has been taken into account. }
\label{fig:radex1}
\end{center}
\end{figure}

\begin{figure}
\begin{center}
\includegraphics[angle=-90,width=8.4cm]{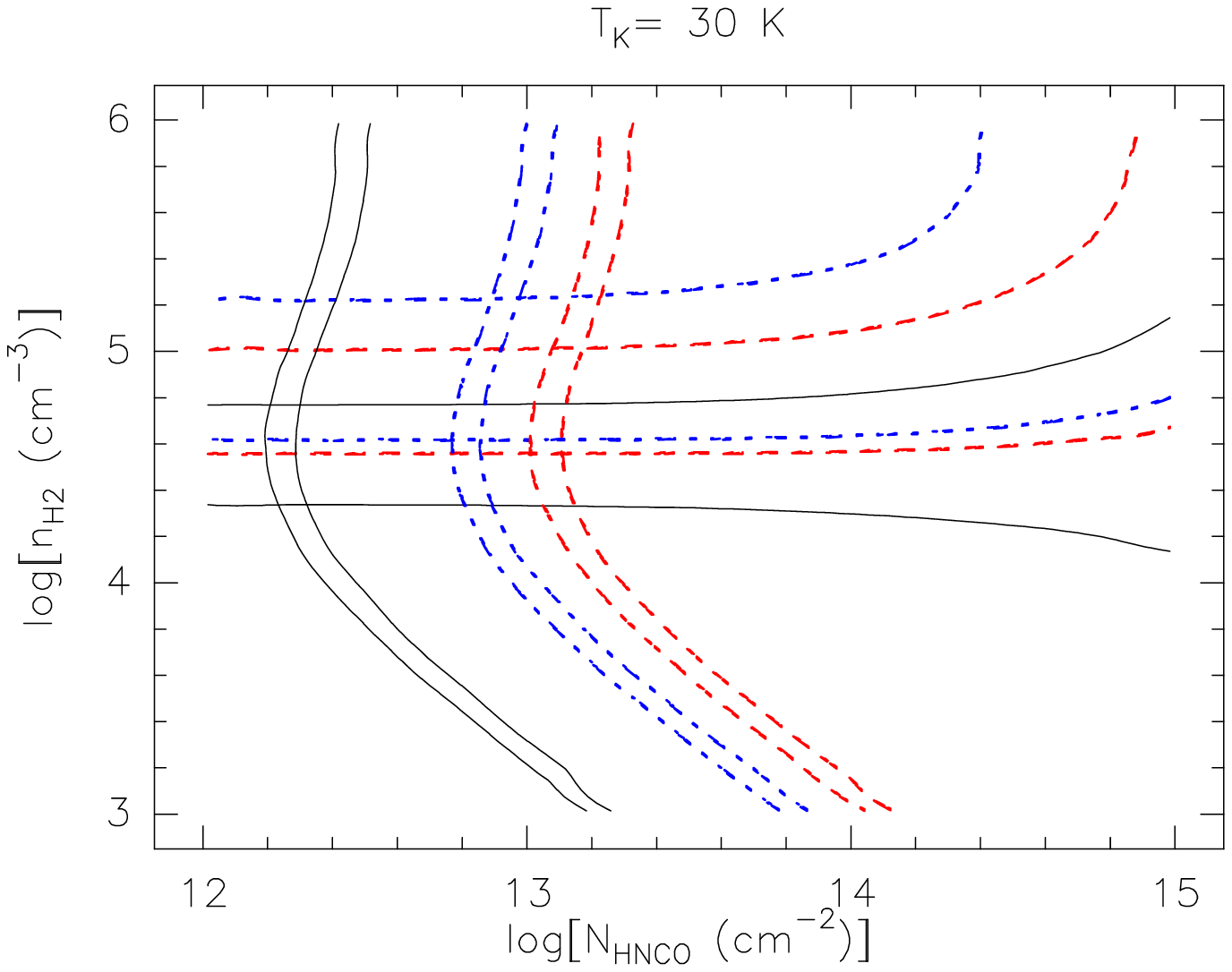}\\
\vspace{2mm}
\includegraphics[angle=-90,width=8.4cm]{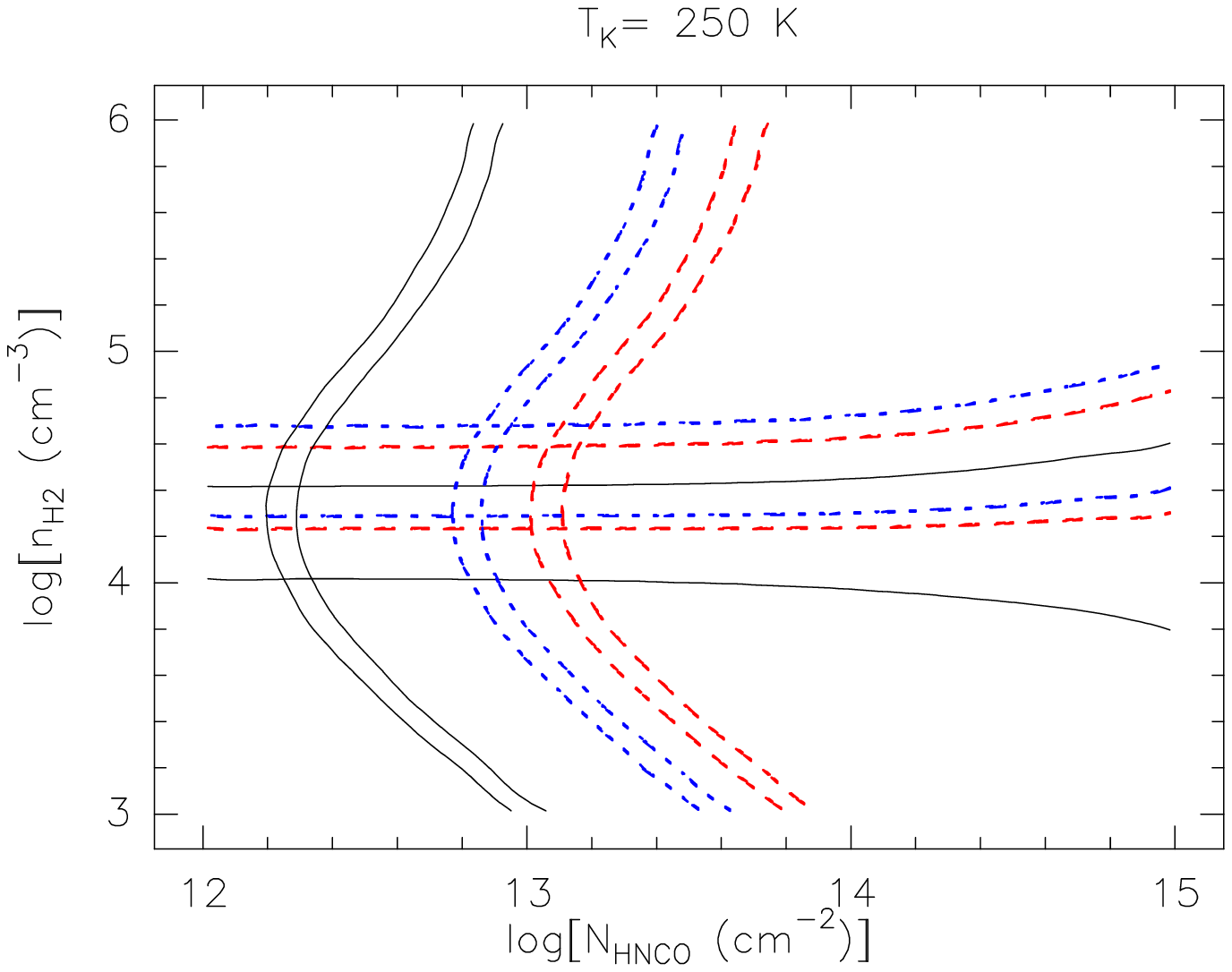}
\caption{Same than  Fig. \ref{fig:radex1} but using $T_{mb}$.}
\label{fig:radex2}
\end{center}
\end{figure}

We have based our analysis on the two lowest lines, which are
detected towards the three observed positions. 
Figure \ref{fig:radex1}  
shows the model predictions for the \lcc \ to \lct\ line ratio and
the \lct \ brightness temperature (for a source size of 10")
as a function of $n_{H_{2}}$ and $N_{HNCO}$.
The same diagrams  for a source size equal to the
telescope main beam (using $T_{mb}$) are shown in Fig.~\ref{fig:radex2}. 

The kinetic temperature cannot be constrained since 
equally good solutions can be found just changing the hydrogen density by $\sim 0.5$~dex.
For a given source size and kinetic temperature, the H$_2$ density  is similar for B1 and B2.
The density varies from $10^{4}-10^{4.4}$~\cmmt \ for a source size of 10" and $T_K=250$ K, to
$10^{4.6}-10^{5.2}$~\cmmt \  for sources filling the main beam and  $T_K=30$ K.
In L1157-mm, $n_{H_{2}}$ is lower than that in B1 and B2 by $0.2-0.3$~dex.

The total HNCO columns densities are 0.2-1.3, 0.6-2.5 and 1.3-5.0
(in units of $10^{13}$ \cmmd) for mm, B1, and B2, respectively (Table~\ref{tab:results}).
The lower limits are similar to those derived using LTE,
while the upper limits for B1 and B2 are somewhat higher because opacity
effects start playing a role.
Table~\ref{tab:results} also shows HNCO abundances computed with the {\sc radex}
estimations of the HNCO column density and the total H$_2$ column density
derived from CO by \cite{Bachiller97}.
The HNCO abundances are
$(0.3-1.2)\times10^{-9}$ for L1157-mm, $(4.3-17.9)\times10^{-9}$ for L1157-B1 and 
$(25-96)\times 10^{-9}$ for L1157-B2 (Table~\ref{tab:results}).
Therefore, there is a clear increase of the HNCO abundance in the shocked regions,
by a factor of 6-14 in B1 and by a factor of 34-83 in B2.
{\it This is the first direct evidence that HNCO is indeed a good tracer of shocked gas}.

In the range of physical conditions and  HNCO column densities
derived from the  \lcc \ and \lct\  lines,
the model predictions for the \lsc \ to \lcc  \ ratio is 0.6-0.7.
In contrast, the measured ratio is 0.3-0.4 for L1157-mm,
and it is lower than 0.3 for B1 and B2.
{As already mentioned above when discussing the
LTE analysis, a multi-transition survey would be needed to fully
understand the excitation of the molecule.
In addition, HNCO maps with high angular resolution  
will also be needed to determine the positions
and the sizes of the different HNCO clumps,
in particular to see if there could be some HNCO
emission out of the IRAM-30m beam at 2mm. 
}

\subsubsection{Comparison with other Galactic sources}

Table \ref{tab:comp} shows  HNCO abundances  with respect to H$_2$
 measured  other Galactic sources as  translucent clouds, photon-dominated regions (PDRs), dense cores and hot cores.
The highest HNCO abundance  is found in hot cores   and dense cores:
up to $\sim 8.7\,10^{-9}$.
The HNCO abundance in L1157-mm is similar to that measured in 
those dense cores and hot cores with the lowest HNCO abundance
and in the only ``hot corino" where the molecule has been previously
 detected, IRAS 16293  \citep{vanDishoeck95}.
The HNCO abundance in the L1157-B1 shock is 
comparable, but can even be higher than that in hot cores and dense cores.
Regarding the L1157-B2 shock, the HNCO abundance is at least 4-11
times higher than that in hot cores and dense cores.
{\it The HNCO abundance in the L1157-B2 shock
 are indeed the highest HNCO abundances
ever measured.} 

\begin{table}
\caption{Measured fluxes}
\label{tab:lines}
\centering
\begin{tabular}{llll}
\hline
\hline
  Line             &  mm\tablefootmark{a} &  B1\tablefootmark{b} &  
  B2\tablefootmark{c} \\
               & K \kms & K \kms & K \kms \\
\hline
 $4_{0,4}-3_{0,3}$  &  0.19      & 0.71         &  1.24    \\
$5_{0,5}-4_{0,4}$  &  0.19      & 0.84         & 1.38    \\
$6_{0,6}-5_{0,5}$  &  0.073      & $< 0.25$ & $< 0.20$  \\
\hline
\end{tabular}
\tablefoot{ Computed in the velocity range from:
\tablefoottext{a} 1.1 to 3.7 km\,s$^{-1}$, \tablefoottext{b}  $-4.8$ to 5.0 km\,s$^{-1}$, 
\tablefoottext{c}$-2.6$ to 6.1 km\,s$^{-1}$.\\
}
\end{table}

\begin{table}
\caption{Results}
\label{tab:results}
\centering
\begin{tabular}{lllll}
\hline
\hline
Source & T$_{ex}^{54}$\tablefootmark{a} & N$_{HNCO}$\tablefootmark{b} & N$_{HNCO}$\tablefootmark{c} & X(HNCO)\tablefootmark{d} \\
       &  K  & 10$^{13}$ \cmmd & 10$^{13}$ \cmmd & 10$^{-9}$ \\
\hline
mm     & 8-11          & 0.2-1.3    & 0.2-1.3  & 0.3-1.2 \\
B1     & 11-19         & 0.6-2.5    & 0.6-2.5  & 4.3-17.9 \\
B2     & 9-16          & 1.3-5.0    & 1.3-5.0  & 25-96   \\
\hline
\end{tabular}
\tablefoot{ 
\tablefoottext{a} Excitation temperatures derived from the populations  in the 
$5_{0,5}$ and $4_{0,4}$ levels.
\tablefoottext{b} HNCO column density assuming LTE, optically thin lines and 
extrapolating using T$_{ex}^{54}$.
\tablefoottext{c} HNCO column density computed  using {\sc radex}.
\tablefoottext{d} HNCO abundance with respect to H$_2$  \citep[using the 
H$_2$ column densities given by][]{Bachiller97}
}
\end{table}

\section{Discussion}

\subsection{HNCO formation in molecular clouds and hot cores}
\label{sect:chemistryhnco}
The possible formation pathways of HNCO in shocks have never been modeled.
In contrast, HNCO has been included in some models of dark clouds and hot core chemistry.
In the models by \cite{Iglesias77}, HNCO is produced by the ion-neutral reaction
 $\mathrm{H}_2 + \mathrm{HNCO}^+ \rightarrow \mathrm{H}_2 \mathrm{NCO}^+ + \mathrm{H} $
followed by
 $\mathrm{H}_2\mathrm{NCO}^+ + e^- \rightarrow \mathrm{HNCO} + \mathrm{H}$.
\cite{Turner99} have considered three possible formation pathways, among them the only efficient one is the neutral-neutral reaction 
$\mathrm{CN} + \mathrm{O}_2 \rightarrow \mathrm{NCO} + \mathrm{O}$ 
followed by
$ \mathrm{NCO} + \mathrm{H}_2 \rightarrow \mathrm{HNCO}+\mathrm{H}$.
{  The last reaction
has an activation barrier  that has been estimated to be $\sim 1160~K$ by \cite{Turner99} 
but that could be as high as 4465~K \citep{Tideswell10}.
In any case, this reaction is not efficient 
at the typical temperature of a hot core ($\sim$ 200 K).
Therefore, 
gas phase chemistry alone cannot explain the HNCO abundances in hot cores.
This is also the case for other species as the complex organic molecules.
Chemical models have consequently been developed to include 
reactions on the dust grains surface
\citep{Caselli93, Garrod08, Tideswell10}. 
In those works, hot cores are modeled in two stages.

In the first phase, a dark cloud of $n_H \sim 10^4$~\cmmt \  suffers an isothermal collapse.
The collapse phase is halted once $n_H \sim 10^7$ \cmmt , which occurs 
after approximately $5\,10^{5}$ years (free-fall time). 
The typical temperature at this phase is 10-20 K and surface chemistry is very important then.
Surface chemistry models \citep{Garrod08, Tideswell10} produce
HNCO as a secondary radical formed via the reaction
 $\mathrm{NH+CO} \rightarrow \mathrm{HNCO}$.
\footnote{HNCO formation on the grain surfaces could also occur by 
reactions of OCN$^-$ with NH$_4^+$ or H$_3$O$^+$. 
The inverse reactions (HNCO with NH$_3$ or H$_2$O) 
are invoked to explain the presence of OCN$^-$ ices in the dust grains, 
\citep{Demyk98, vanBroekhuizen04}.
To our knowledge, these reactions have never
 been included in chemical models.}
On the grain surface,  
HNCO can reach a maximum abundance 
of 10$^{-5}$ \citep{Tideswell10} before being
destroyed by new reactions with primary radicals (H, CH$_3$, HCO, NH, ...)
giving complex species as HNCHO, HNCOCHO, or CH$_3$CONH \citep{Garrod08}.

The cloud-collapse phase is followed by a warm-up phase in which
the temperature increases to $\sim 200$~K.
The timescale of the warm-up phase can be as short as $\sim 5\,10^4$~yr in hot cores
but it can reach 10$^6$ yr in their low mass equivalent ({\it hot corinos}), 
giving a similar although  slightly different chemistry \citep{Garrod08}.

During the warm-up phase molecules are evaporated  from the grain mantles.
With respect to HNCO, the models by \cite{Garrod08}  and \cite{Tideswell10} show
that its abundance increases monotonously  with time  in the gas phase,
which implies  that HNCO is not directly ejected from the grains.
Instead, HNCO is formed by the destruction in the gas phase of the complex molecules
formed from HNCO on the grain surface \citep{Garrod08, Tideswell10}.
This mechanism can explain the HNCO abundances measured in hot cores of $10^{-9}-10^{-8}$.
As already mentioned, the contribution of the \cite{Turner99} and \cite{Iglesias77}
formation pathways is negligible at the moderate temperature of a hot core.

The abundance of HNCO in L1157-mm is similar to that measured in the hot
corino of the only other low mass protostar where HNCO has been observed 
(IRAS16293-2422, see Table \ref{tab:comp}).
Therefore, 
the HNCO abundance measured in L1157-mm can also  be explained in the context of the
hot-cores models discussed above. 
This would imply that L1157-mm can be considered as a hot corino, which
is indeed in agreement with the intense emission of water detected by Herschel 
towards mm (Nisini et al. {\it in prep}). }
However, more observations will be needed to clearly establish 
the hot corino character of L1157-mm, in particular observations of
complex organic molecules.

\subsection{HNCO formation in shocks}
\label{sect:chemistry}
What is the formation pathway of HNCO in the L1157 outflow shocks? 
The most likely scenario is a combination of grain surface and gas phase chemistry. 
{ In the context of shocks, molecules formed on the grain surfaces can be ejected to the
gas phase due to grain sputtering instead of thermal evaporation as in hot cores.
Therefore, the chemistry will be sensitive to the grain mantles composition at
the time of the shock arrival.
Currently, there are no models that study the HNCO abundance in dark clouds
including grain surface chemistry. 
Dark clouds models by \cite{Tideswell10} only consider  gas phase reactions.
Therefore, the exact grain mantle composition before the
 arrival of the shock in L1157-B1 and L1157-B2 is not known.
Nevertheless, one can compare with the cloud-collapse phase of hot core models 
(the main difference is that the density during the collapse phase reaches
higher values than in dark clouds). 
As discussed in the previous section,
during the collapse phase
the HNCO abundance on the grain surfaces has a peak of 10$^{-5}$,
grain sputtering at that 
time will immediately give rise to very high HNCO abundance in gas phase.  
Even the very high HNCO abundance in L1157-B2 ($\sim 10^{-7}$) could be accounted
for in this context. 

On the other hand, gas phase chemistry in shocks differs considerably  from
that in hot cores because the temperature in the shocked gas can be much 
higher than that in a hot core.
As already pointed out by \cite{Zinchenko00},
in shocks  the reaction 
$\mathrm{NCO} + \mathrm{H}_2 $  can be efficient in spite of  the activation barrier.
In addition, the reaction
$\mathrm{CN} + \mathrm{O}_2 \rightarrow \mathrm{NCO} + \mathrm{O}$ will be favored
by the high O$_2$ abundances that are predicted in post-shock gas
\citep{Bergin98, Gusdorf08a, Gusdorf08b}. 
\footnote{This does not mean that the \textit{current} $O_2$ abundance in the 
L 1157 shocks is high, but it may have been
high just after the passage of a shock-wave as predicted by shock models.
Since the chemistry in shocked regions is fast, O$_2$
can be rapidly converted into other species.
Therefore, there is not any contradiction with the fact that 
the O$_2$ abundance in the interstellar medium has been found to be low
\citep[$< 10^{-7}$][]{Larsson07}.
 }

Using \cite{Tideswell10} results, it is possible to  verify if 
gas phase chemistry alone could explain the HNCO abundance measured in
the L1157 shocks.
In one  of their dark cloud models (DC1), they have
studied the HNCO formation exclusively via the gas phase neutral-neutral reactions
 of \cite{Turner99} as a function of the rate coefficient for the reaction 
$ \mathrm{NCO} + \mathrm{H}_2 \rightarrow \mathrm{HNCO}+\mathrm{H}$.
Their Fig. 1 shows that a rate coefficient $k$ higher than
$10^{-14}$ cm$^{3}$\,s$^{-1}$ is needed to explain the observed abundances in L1157 
of $10^{-8}-10^{-7}$.
However, to have high HNCO abundance (a few $10^{-8}$) {\it in a short time}
(several 10$^3$ yr) the rate coefficient must be higher than $10^{-12}$~cm$^{3}$\,s$^{-1}$.
Taking into account the formula $k(T)=1.5\,10^{-11} \, \exp(-4465/T) \, \mathrm{cm}^{-3} \, \mathrm{s}^{-1}$ given by \cite{Tideswell10}, $k=10^{-12}$~cm$^{3}$\,s$^{-1}$
implies a temperature higher than 1600 K. 
In contrast, the minimum temperature would be 428 K if the
 activation barrier is only  1160~K \citep{Turner99}.
\cite{Gusdorf08b} have recently computed shock models to compare  the modelled
SiO and H$_2$ emission to observations of L1157. 
The best agreement between models and observations is found 
for a preshock density of 10$^{4}$ cm$^{-3}$ and a shock speed of 20 km\,s$^{-1}$.
In such a shock, the gas temperature is higher than 430-1600 K
for a few thousand years, which is consistent with the dynamical age of the B1 and B2 shocks \citep[2000-3000 yr, ][]{Gueth98}
and with the chemical time necessary 
to have an HNCO abundance of a few  10$^{-8}$
only with gas phase reactions \citep[Figure 1 of ][]{Tideswell10}. 
Therefore, gas phase chemistry alone could explain the HNCO abundance in L1157-B1.
However, it is unclear whether a HNCO abundance as high as that 
measured in L1157-B2 ($\sim 10^{-7}$) can be attained 
in a few thousand years
only with gas phase reactions. }

Therefore, we reckon that the most likely explanation to the
high HNCO abundances in L1157 is 
dust grain mantles processing by the shock waves followed by 
 neutral-neutral reactions in gas phase. 

\begin{table}
\caption{HNCO abundance with respect to H$_2$ in different astronomical environments.}
\label{tab:comp}
\centering
\begin{tabular}{lll}
\hline
Source        & X(HNCO)   & References\\  
              & 10$^{-9}$ &\\
\hline
L1157-B2               & 25-96      &1\\
Translucent clouds & 0.2-5      &2  \\  
Orion bar PDR      & $<10^{-2}$ &3 \\
Dense cores       & 0.2-8.7    &4  \\
Orion Hot core     & 5          &5\\
SgrB2M             & 1          &6\\
W3(H$_2$O)         & 5.0-27     &7    \\
G34.3+0.15        & 1.4        &8 \\
IRAS 16293        & 0.17-4.0   & 9, 10   \\
GC clouds         & 3-48       & 11,12\\
Starburst galaxies & 1-6.3      & 13\\
\hline
\end{tabular}\\
\tablebib{(1)~This work; (2)~\citet{Turner99};(3)~\citet{Jansen95}; 
(4)~\citet{Zinchenko00}; (5)~\citet{Blake87}; (6)~\citet{Minh98};
(7)~\citet{Helmich97}; (8)~\citet{Bockelee-Morvan00}, using the HNCO  column density given by \cite{MacDonald96} and the H$_2$ column density given by \cite{Hatchell98}; 
(9)~\citet{vanDishoeck95}; (10)~\citet{Bisschop08};
 (11)~\citet{Rizzo00}; (12)~\citet{Martin08}, and references within; 
(13)~\citet{Martin09} 
}
\end{table}

\subsection{Chemical differences between B1 and B2 and the role of the O$_2$ chemistry}
\label{sect:o2chemistry}

We have discussed the line profiles and the line intensities of different 
species in Sect.~\ref{sect:profiles}.
The HNCO profiles are very similar to those of SO and SO$_2$.
In addition, the lines of most of the molecules show similar intensities
in B1 and B2 with the exception of CN, which is more intense in B1, and
HNCO, SO$_2$ and OCS, whose lines are 
more intense emission in B2 than in B1.
These similarities of HNCO and the sulfured molecules is somewhat puzzling.

The actual reason of the  chemical differences in B1 and B2 is not clear.
The present gas density in B1 and B2 is similar \citep[see Figs. \ref{fig:radex1} and \ref{fig:radex2} and][]{Bachiller97, Nisini07}. 
 In contrast, based on the line
profiles, the shock velocity can be higher in B1 than in B2.
Taking into account the extreme SiO line wings, the difference could reach
10 \kms, which after the \cite{Gusdorf08a} models could be significant.

Alternatively, \cite{Bachiller97} have suggested that the chemical
 differences between B1 and B2 could be due to different shock ages.
Indeed, in contrast to SO$_2$ and OCS,
the H$_2$S abundance is lower in B2 than in B1.
This is consistent since H$_2$S is a parent molecule for other
sulfured species.  
H$_2$S is formed in the grain
surfaces and  released to the gas
phase by effect of the shock waves.  
Other sulfur-bearing molecules like
SO and SO$_2$ are produced in gas phase very quickly (few 10$^3$ yr) 
via reactions with H, OH and O$_2$ \citep{Pineau93, Wakelam05}.
One possible explanation of the differences in B1 and B2 is that the
B2 shock could be older than that in B1
and that H$_2$S has been converted into SO and SO$_2$.
{Even if there are uncertainties regarding the sulfur chemistry
\citep[see for instance][and references therein]{Codella05}, sulfur-bearing
molecules like H$_2$S, SO, SO$_2$, H$_2$CS and OCS have been invoked to be
 potential valuable 
probes of chemical evolution \citep{Codella05, Wakelam05, Herpin09}.
In particular, \cite{Wakelam05} have found that the   SO$_2$/SO ratio 
 increases with the shock age. } 
 The higher SO$_2$/SO ratio in B2 compared to
 B1 also points to an older shock in B2.
\cite{Bachiller97} have also
proposed that the different CN abundance in B1 and B2 
can  be accounted for in
this scenario of an older shock in B2,
since after an enhancement of CN in the shock, reactions like
CN+O $\rightarrow$ CO+N  could be very efficient to destroy CN.
Our data globally agree with this scenario, nevertheless 
the clear anticorrelation of the CN and HNCO intensities and the 
similar abundances of CN and HNCO in L1157-B1 suggest that
$all$ the CN can indeed be transformed into HNCO once the shock
has increased the temperature and the neutral-neutral reactions
form NCO from CN and O$_2$ and HNCO from NCO (see Sect. \ref{sect:chemistry}).
Therefore, the chemical differences from B1 and B2 do support
the scenario proposed in Sect. \ref{sect:chemistry} to explain the
HNCO abundances in the L1157 shocks.

In addition, the scheme proposed in Sect. \ref{sect:chemistry}
also gives  insight on the possible 
 link of HNCO and the sulfured molecules, in particular 
with SO and SO$_2$.
In shocks these molecules can be produced by the reactions 
$\mathrm{S+O_2}  \rightarrow \mathrm{SO+O}$ and 
$\mathrm{SO+OH}\rightarrow \mathrm{SO_2+OH}$
\citep{Pineau93,  Charnley97, Wakelam05}.
Therefore, the common link with HNCO would be   formation pathways 
involving  O$_2$. 

Another important shock tracer that 
could be linked to the O$_2$ chemistry is SiO.
Nevertheless, the situation regarding this molecule is more complex.
First, recent  models  can explain the SiO emission in shocks 
without a significant contribution of Si oxidation in gas phase, 
either by sputtering in gas-grain collisions 
if there is already SiO in the grain mantles \citep[][]{Gusdorf08b} 
or by dust vaporization in grain-grain collisions 
\citep{Guillet09}.
Second, if one  assumes that SiO is formed in gas phase, there is
a threshold shock velocity of $\sim 25$ \kms \ in order to
eject Si from the grain cores by sputtering \citep{Gusdorf08a}.
In contrast, other molecules as the organics are in the grain mantles
and there is not such a high velocity threshold for them to be
ejected to gas phase.
{ Therefore, both if SiO comes directly from the grains \citep{Gusdorf08b, Guillet09}
or if it is formed in gas phase \citep{Gusdorf08a}, due to the shock velocity
threshold, it is not expected to find similarities (and not observed) in the SiO emission and that of SO, SO$_2$ and HNCO.

 It is interesting to note that the impressive SiO blue wing in B1 \citep{Bachiller97}
suggests that SiO formation is indeed favored, in comparison to other molecules,
in the higher velocity gas. 
In this context, chemical differences in B1 and B2 could also be due to
different shock velocities. 
If the shock velocity in B2 is actually lower than in B1 and the 
SiO formation less efficient, more O$_2$ will remain available in B2
to form HNCO, SO and SO$_2$.} 



\subsection{On the  origin of the HNCO emission in galactic nuclei}
The HNCO abundance in the molecular clouds in the center of the Milky Way
and in the nuclei of starburst galaxies is similar or a bit higher (at most 
by a factor of 2) than in Galactic hot cores (Table \ref{tab:comp}).
HNCO becomes another piece in the well known puzzle of
the chemistry of  the Galactic center molecular clouds.
This puzzle can be summarized as follows:
the abundance of SiO and complex organic molecules in the Galacic center
is as high as in hot cores 
\citep{Martin-Pintado97, Rodriguez06, Requena06}.
In contrast, ({\it i}) the emission in the Galactic center 
is extended over the central 300 pc and it does not resemble 
a collection of discrete sources with the size of a hot core ($\sim$ 0.1 pc),
({\it ii}) the gas density ($10^3-10^4$ \cmmt) 
is  much lower than in hot cores ({\it iii}) and the dust temperatures ($<30$ K) are also much lower than in hot cores.
The Galactic center clouds present a ``hot core chemistry without hot cores"
\citep{Requena06}.
The origin of this chemistry is not known although it is thought to be due to
some type of mechanical processes as shock waves 
\citep{Martin-Pintado97, Martin-Pintado01, Rodriguez04}.
The origin of the shocks can be related to the complex dynamics in
the inner regions of the Galaxy 
\citep{Huttemeister98, Rodriguez06, Rodriguez08}.

Based on the spatial distribution of the HNCO and the comparison with other 
species, as CH$_3$OH, \cite{Meier05} have also suggested that the HNCO emission in
IC342 is tracing shocks. 
Recently, \cite{Martin08, Martin09} have also proposed that shocks could be the 
explanation of the  high HNCO abundances measured in galactic nuclei.

Discussing the precise origin of shocks in Galactic nuclei is out of the 
scope of this paper.
Nevertheless, our observations support the scenario of HNCO tracing shocks 
in galactic nuclei since our L1157 results probe
a medium of moderate H$_2$ density 
where the HNCO abundace is indeed high due to the grain processing and gas heating by shock waves.

\section{Summary}

We have observed three lines of HNCO towards the protostar L1157 and its
associated molecular outflow.
Our aim is to characterize the HNCO properties in shocked gas.
HNCO is well detected in the shocked gas,
where the abundance increases by a factor of 6-34
with respect to the abundance in the protostar L1157-mm.
The  abundance in B1 and B2 is 0.4-1.8 \,10$^{-8}$ 
and 0.3-1 \,10$^{-7}$, respectively.
The abundance in B2 is the
highest ever measured, considerably higher than those in hot cores 
(a few $10^{-9}$) and galactic nuclei ($10^{-9}-10^{-8}$).
Our results probe that the HNCO abundances measured in galactic nuclei
can easily be attained in shocked gas, providing a solid
basis to previous suggestions that the extended HNCO in galactic 
nuclei could trace large scale shocks \citep{Meier05,Minh06}.

The dominant formation pathway of HNCO in hot cores
is grain mantle evaporation of complex molecules formed from HNCO and
subsequent dissociation to give again HNCO.
In addition, there is contribution from gas phase reactions, but it is minor
 due to the high activation barriers of some reactions
\citep{Tideswell10}.
In the L1157 shocks we propose a relatively similar formation pathway 
although with different relative importance of the dust surface and gas phase
chemistry.
First there should be  grain mantles erosion by the shock waves that will
increase the HNCO abundance in gas phase.
Second, in contrast to hot cores, the neutral-neutral reactions in gas phase
can be the dominant formation pathway due 
to the  higher gas temperatures in the shocked gas 
with respect to the typical  temperature of hot cores.
These reactions start with the formation  of OCN from CN and O$_2$,
which is expected to be very abundant in gas phase in a magneto-hydrodynamic 
shock of relatively low velocity (20-30 \kms) like the L1157 shocks
\citep[][]{Gusdorf08a}.
The observed anticorrelation of CN and HNCO fluxes also support this scenario.

HNCO line profiles are very similar to those of CH$_3$OH and H$_2$CO lines, 
which is not surprising since all those molecules have, at least partially, a common
origin linked to the grain surface chemistry and subsequent desorption.
In addition, HNCO line profiles are also similar to some sulfur-bearing molecules
like SO and SO$_2$.
Another common property of HNCO and the sulfur-bearing
molecules is that they are the only species that exhibit a more intense emission
in B2 than B1.
We propose that the common link of HNCO and the sulfured molecules 
is that in both cases there are formation pathways with reactions involving O$_2$.
HNCO together with the sulfured molecules are
good probes of chemical differentiation in shocks. 
In particular, in the case of 
L1157 the highest abundances in B2 than in B1
could be due to a lower shock velocity and/or an older shock
in B2 than in B1.

\begin{acknowledgements}
NJR-F acknowledges fruitful discussions on chemistry and molecular outflows 
with S.~Cabrit, A.~Fuente and C.~Codella.
This work has been partially funded by the grant
ANR-09-BLAN-0231-01 from the French {\it Agence Nationale de la Recherche} as part
of the SCHISM project.
We thank an anonymous referee for a very detailed report 
that  helped to correct and clarify a number of issues in the 
original version of the manuscript. 
\end{acknowledgements}

\bibliographystyle{aa}
\bibliography{paper_HNCO_L1157}

\end{document}